\documentclass[accepted]{article}

\usepackage{microtype}
\usepackage{graphicx}
\usepackage{subfigure}
\usepackage{booktabs} 

\usepackage{array}
\newcolumntype{L}[1]{>{\raggedright\arraybackslash}p{#1}} 
\newcolumntype{C}[1]{>{\centering\arraybackslash}p{#1}}   
\newcolumntype{Y}{>{\RaggedRight\arraybackslash}X}        
\usepackage{tabularx,multirow,ragged2e}

\usepackage{hyperref}
\usepackage{xurl}
\hypersetup{breaklinks=true}

\usepackage{mlsys2025}

\mlsystitlerunning{PIM-SHERPA: Software Method for On-device LLM Inference by Resolving PIM Memory Attribute and Layout Inconsistencies}

\makeatletter
\renewcommand{\mlsys@appearing}{}
\makeatother
\begin{document}

\twocolumn[
\mlsystitle{PIM-SHERPA: Software Method for On-device LLM Inference by Resolving PIM Memory Attribute and Layout Inconsistencies}




\begin{mlsysauthorlist}
\mlsysauthor{Sunjung Lee}{sait}
\mlsysauthor{Sanghoon Cha}{sait}
\mlsysauthor{Hyeonsu Kim}{sait}
\mlsysauthor{Seungwoo Seo}{sait}
\mlsysauthor{Yuhwan Ro}{sait}
\mlsysauthor{Sukhan Lee}{sec}
\mlsysauthor{Byeongho Kim}{sec}
\mlsysauthor{Yongjun Park}{yonsei}
\mlsysauthor{Kyomin Sohn}{sec}
\mlsysauthor{Seungwon Lee}{sait}
\mlsysauthor{Jaehoon Yu}{sait}

\end{mlsysauthorlist}

\mlsysaffiliation{sait}{Samsung Advanced Institute of Technology, Suwon, South Korea}
\mlsysaffiliation{sec}{Samsung Electronics, Hwaseong, South Korea}
\mlsysaffiliation{yonsei}{Yonsei University, Seoul, South Korea}

\mlsyscorrespondingauthor{Jaehoon Yu}{jae-hoon.yu@samsung.com}

\mlsyskeywords{Machine Learning, MLSys}

\vskip 0.3in

\begin{abstract}
On-device deployments of large language models (LLMs) are rapidly proliferating across mobile and edge platforms. LLM inference comprises a compute-intensive prefill phase and a memory bandwidth-intensive decode phase, and the decode phase has been widely recognized as well-suited to processing-in-memory (PIM) in both academia and industry. However, practical PIM-enabled systems face two obstacles between these phases, a memory attribute inconsistency in which prefill favors placing weights in a cacheable region for reuse whereas decode requires weights in a non-cacheable region to reliably trigger PIM, and a weight layout inconsistency between host-friendly and PIM-aware layouts. To address these problems, we introduce \textit{PIM-SHERPA}, a software-only method for efficient on-device LLM inference by resolving PIM memory attribute and layout inconsistencies. PIM-SHERPA provides two approaches, DRAM double buffering (DDB), which keeps a single PIM-aware weights in the non-cacheable region while prefetching the swizzled weights of the next layer into small cacheable buffers, and online weight rearrangement with swizzled memory copy (OWR), which performs the on-demand swizzled memory copy immediately before GEMM. Compared to a baseline PIM emulation system, PIM-SHERPA achieves approximately 47.8 - 49.7\% memory capacity savings while maintaining comparable performance to the theoretical maximum on the Llama 3.2 model. To the best of our knowledge, this is the first work to identify the memory attribute inconsistency and propose effective solutions on product-level PIM-enabled systems.
\end{abstract}
]



\printAffiliationsAndNotice{}  

\section{Introduction}
\label{sec:introduction}

Large language models (LLMs) have rapidly advanced language understanding and generation, driving applications such as conversational assistants and real-time translation. As these services proliferate, there is strong demand to run inference on devices such as smartphones and edge systems to eliminate network latency and preserve user privacy. Major vendors have already offered LLM features on mobile platforms ~\cite{samsung-ai, google-ai, apple-ai, qualcomm-ai}, and lightweight runtimes are emerging to support such deployments~\cite{executorch, tinychatengine}.

LLM inference consists of the prefill and decode phases, which have distinct characteristics. In a single-batch scenario~\cite{yu2024cambricon}, prefill processes many tokens, so the linear layers execute in the form of matrix-matrix multiplication (GEMM). Because prefill is compute-intensive, it runs best on a GPU, NPU, or CPU. Decode processes a single token at a time, which turns the linear layers into matrix-vector multiplication (GEMV) and shifts the bottleneck to memory bandwidth. To alleviate the DRAM bandwidth bottleneck in on-device environments, processing-in-memory (PIM) architectures, specifically LPDDR-PIM, have recently emerged as a promising alternative~\cite{kim2020mvid, kim2024breakthrough, ibrahim2024pimnast, jeun2025foldpim, he2025lpspec} by utilizing up to eight times higher effective DRAM bandwidth.

However, applying LPDDR-PIM to on-device inference introduces critical system-level challenges  between the prefill and decode phases: 1) \textbf{memory attribute inconsistency} and 2) \textbf{weight layout inconsistency}.
The memory attribute inconsistency arises because the weight requires contradictory memory attributes between phases. Prefill requires weights in a cacheable region to maximize the cache reuse opportunities. Decode requires the weights in a non-cacheable region to utilize PIM. More specifically, DRAM requests trigger PIM execution. If the weights are in a cacheable region, cache hits may prevent requests from reaching the memory controller, which blocks PIM execution. The weight layout inconsistency is due to the different optimal memory layout for each phase. Host-side operation prefers a host-friendly contiguous layout that benefits from channel and bank interleaving. In contrast, PIM operation prefers a PIM-aware layout that places data in contiguous DRAM columns within each DRAM bank to maximize in-bank SIMD utilization~\cite{kim2024breakthrough, ibrahim2024pimnast}. These conflicting data layout requirements arise when LLM inference performs both prefill and decode processes.

Prior studies on PIM-based acceleration of LLM inference~\cite{ibrahim2024pimnast, kim2024skhynix_aimx, seo2024ianus, park2024attacc, heo2024neupims} have not explicitly addressed the memory attribute and weight layout inconsistencies.
HBM-PIM~\cite{kim2024breakthrough} simply duplicates weights by keeping a cacheable host-friendly copy and a non-cacheable PIM-aware copy, which nearly doubles memory footprint and narrows the set of models that can be run by on-device. FACIL~\cite{seo2025facil} augments the memory controller to resolve the layout inconsistency so that both phases can access suitable layouts from a single PIM-aware weights, but the approach requires hardware changes and leaves the memory attribute issue unresolved. 

To address these challenges, we introduce a \underline{\textbf{s}}oftware met\underline{\textbf{h}}od for efficient on-device LLM inf\underline{\textbf{er}}ence by resolving \underline{\textbf{P}}IM memory \underline{\textbf{a}}ttribute and layout inconsistencies, named \textbf{PIM-SHERPA}. We propose two software-only methods tailored to the trend toward longer input sequence lengths driven by interactive LLMs~\cite{jeong2025flashgen} and on-device retrieval-augmented generation (RAG) ~\cite{google-ai-edge, seemakhupt2024edgerag, park2025mobilerag}. 
DRAM double buffering (DDB) reserves a small DRAM buffer in the cacheable region, while the current layer runs GEMM, and prefetches weights for the next layer into the cacheable buffer. By fetching weights in parallel with computation, the rearrangement overhead is overlapped with compute latency. A swizzled memory copy transforms weights from the PIM-aware layout to a host-friendly layout and stores them in the cacheable region. Online weight rearrangement with swizzled memory copy (OWR) assumes sufficiently long input sequences and copies the required weights immediately before computation of each layer. Different from DDB, OWR does not require sophisticated synchronization or load-balancing control between copy and computation, which lowers implementation complexity. Both techniques additionally require only small cacheable buffers approximately equal to the weight size of one or two layers, so the memory capacity overhead is minimal compared to weight duplication.

The contributions of our study are as follows:
\begin{itemize}
    \item We identify the memory attribute inconsistency between prefill and decode in PIM-enabled LLM inference. The system-level challenge has not been addressed by the prior works.
    \item We characterize the weight layout inconsistency between host-friendly and PIM-aware models and quantify how na\"{\i}ve  duplication inflates DRAM capacity on representative LLMs.
    \item We propose two software-only solutions, DDB and OWR, that employ small cacheable buffers and introduce a swizzled memory copy to resolve these inconsistencies without modifying existing hardware.
\end{itemize}

\section{Background}
\label{sec:background}

\begin{figure}[!tb]
  \center
  \includegraphics[width=0.85\columnwidth]{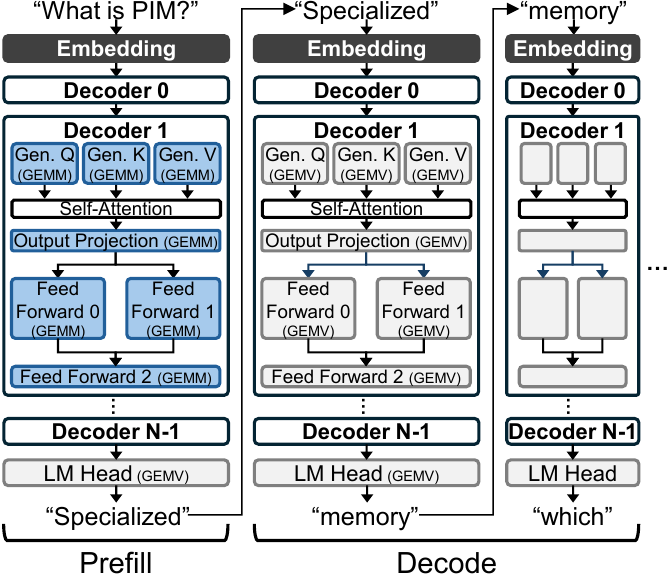}\caption{
    Model structure and inference process of LLM
  }
  \label{fig:llm}
  \vspace{-0.1in}
\end{figure}

\begin{figure}[!tb]
  \center
  \includegraphics[width=0.8\columnwidth]{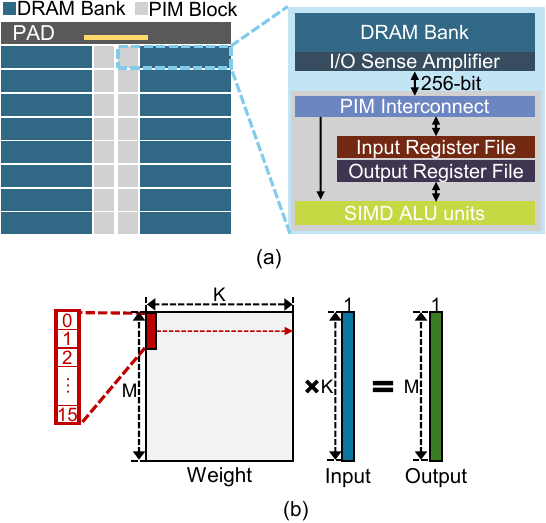}\caption{
    (a) LPDDR-PIM architecture, and (b) GEMV operating sequence of LPDDR-PIM
  }
  \label{fig:pim}
  \vspace{-0.15in}
\end{figure}

\subsection{Large language model inference}
\label{ssec:llm}


LLMs are built on a stack of Transformer decoder layers~\cite{vaswani2017attention}. Figure~\ref{fig:llm} shows the model structure of Llama 3.2~\cite{meta2024llama32_blog}, a representative on-device LLM. The models process embedded input tokens sequentially to compute the probability distribution for the next token. Each layer encompasses self-attention and feed forward networks, typically featuring layer normalization within its sub-layers. LLM inference is divided into two distinct phases: the prefill phase, which handles long input sequences in a single pass and is dominated by compute-intensive GEMM (blue), and the subsequent decode phase, which processes one token per step and is constrained by memory bandwidth due to GEMV (gray)~\cite{he2025waferllm}.

\subsection{LPDDR-PIM architecture}
\label{ssec:pim}

LPDDR-PIM~\cite{kim2024breakthrough} and PIMnast~\cite{ibrahim2024pimnast} embed a SIMD unit inside each PIM block so that a 256-bit burst fetched from a DRAM bank can be processed in a single step. As shown in Figure~\ref{fig:pim}, an LPDDR rank consists of 16 banks, each connected to one or two PIM blocks that include a SIMD unit and input/output register files. In multi-bank PIM mode, each SIMD unit performs 256-bit operations using inputs from local registers and writes results back to output registers. There are multiple data placement strategies for PIM. PIMnast explores data placement strategies for LPDDR-PIM and shows that storing weights in column-major order at the 256-bit request granularity is most efficient (see Figure~\ref{fig:pim}(b)). For FP16 or BF16, a 256-bit burst holds 16 weight elements, each contributing to 16 partial sums during GEMV. This layout reduces host-side accumulation overhead. Our design adopts the placement strategy proposed by PIMnast.

\subsection{Weight rearrangement for PIM}
\label{ssec:rearrangement}

\begin{figure}[!tb]
  \center
  \includegraphics[width=0.89\columnwidth]{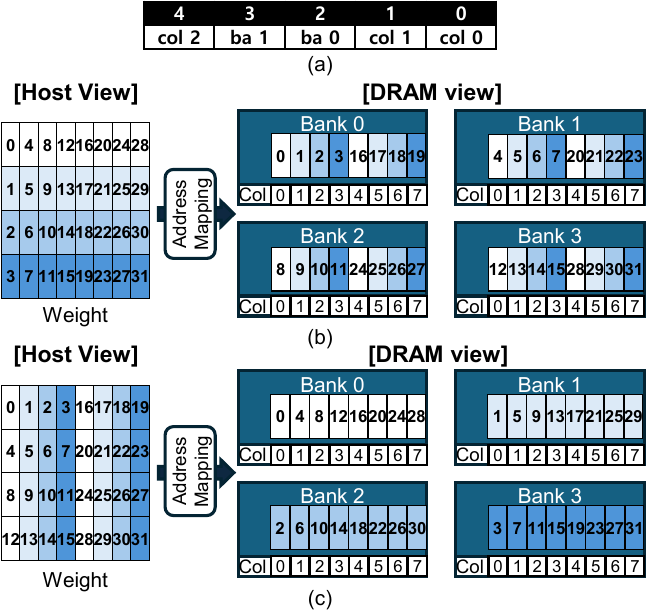}\caption{
    Mapping of host-friendly and PIM-aware weights inside DRAM. (a) Example address map. (b) Host and DRAM view of host-friendly weight and (c) PIM-aware weight where each bank holds a complete matrix row. Each rectangular block represents 16 weight elements in FP16/BF16 data format, which are stored in column-major order.
  }
  \label{fig:replacement}
  \vspace{-0.1in}
\end{figure}

To execute LLMs on a PIM architecture, the weight matrix layout of conventional models must be rearranged to match PIM execution characteristics. Figure~\ref{fig:replacement} illustrates how weights from a conventional model and a PIM-aware model are stored in DRAM. The address map is simplified to one channel, one rank, four banks, one DRAM row, and three DRAM columns (see Figure~\ref{fig:replacement}(a)). Figure~\ref{fig:replacement}(b) shows the conventional host-friendly layout. The host issues sequential addresses along the column-major direction of the weight matrix. To maximize DRAM bandwidth, the weights of sequential addresses are interleaved across multiple channels and banks. In contract, PIM prefers for an entire matrix row to reside within a single bank. As shown in Figure 3(c), placing complete rows per bank at the DRAM level requires a PIM-aware rearrangement of the weights.

\subsection{PIM execution}
\label{ssec:execution-flow}

\begin{algorithm}[tb]
   \caption{HBM-PIM GEMV pseudo code}
   \label{alg:execution-flow}
\begin{algorithmic}
   \FOR{$o = 0$ {\bfseries to} $\textnormal{$NumOutTile$}-1$}
     \FOR{$i = 0$ {\bfseries to} $\textnormal{$NumInputTile$}-1$}
       \STATE \textbf{Write8}(\texttt{InBufPtr}, \texttt{InRfPtr}) 
       \STATE $offset \gets \textbf{AddrGen}(col, ba, ch)$
       \FOR{$r = 0$ {\bfseries to} $\textnormal{$NumTileRow$}-1$}
         \STATE \textbf{Read32}(\texttt{WgtBufPtr} + $offset$)
         \STATE \texttt{WgtBufPtr} $\gets$ \texttt{WgtBufPtr} $+$ \texttt{RowOffset}
       \ENDFOR
       \STATE \texttt{InBufPtr} $\gets$ \texttt{InBufPtr} $+$ \texttt{InTileOffset}
     \ENDFOR
     \STATE \textbf{Read5}(\texttt{DummyAddr}) 
     \STATE \textbf{Write8}(\texttt{OutRfPtr}, \texttt{OutBufPtr})
     \STATE \texttt{OutBufPtr} $\gets$ \texttt{OutBufPtr} $+$ \texttt{OutTileOffset}
   \ENDFOR

\end{algorithmic}
\end{algorithm}

Host controls HBM-PIM~\cite{lee2021pim_isca} through conventional DRAM read and write requests~\cite{pimlibrary}. Algorithm~\ref{alg:execution-flow} illustrates an example of GEMV execution on HBM-PIM as pseudo code. Inputs are first written to input register file via DRAM writes. The host then issues DRAM reads to trigger MAC operations. Because HBM provides 32 DRAM columns per row, reads are grouped to units of 32 for maximizing row-buffer hits. After completing MACs for the entire weight matrix, the host issues additional DRAM reads as NOPs to drain the SIMD ALU pipeline. The results in the output register file are then written back to DRAM.

For correct PIM operation, weight must be mapped to a non-cacheable region. During the MAC phase (\textit{Read32}), each DRAM read corresponds to one MAC operation. If weights are placed in a cacheable region and remain in cache from prior accesses, subsequent reads are served by the cache rather than reaching the memory controller, preventing the PIM execution. 
Moreover, hardware prefetchers or cache locality policies may generate unintended DRAM requests, disrupting synchronization. Therefore, to ensure deterministic execution, all weights must be allocated in non-cacheable region.
\section{Motivation}
\label{sec:motivation}

\subsection{Inconsistency challenges}
\label{ssec:inconsistency}

The prefill and decode phases require different memory attributes and weight layouts for the an LLM weight.
\\
\textbf{Prefill phase (GEMM):} Prefill is compute-intensive and typically executes on the host processor with multiple input tokens. 
During computation, repeated accesses to each weight element by up to the input sequence length (SL) times, leads to high weight reuse opportunity. Therefore, mapping the weight to a cacheable region is proven to be effective because it allows the host to reuse cached data. From a layout perspective, the host prefers a contiguous host-friendly layout, such as column-major or row-major order in the virtual address space.

\textbf{Decode phase (GEMV):} Decode is memory bandwidth-intensive and benefits from PIM when processing a single input token at a time. PIM can exploit up to an eightfold increase in effective DRAM bandwidth during the GEMV operation. PIM is triggered by DRAM requests which are issued by the host. If the weight is mapped to the cacheable region, DRAM requests can hit in cache and cannot reach the memory controller, which prevents the intended PIM operation. Hence, the weight must be mapped to a non-cacheable region. From a layout perspective, PIM requires a PIM-aware layout in which matrix data are contiguous within each DRAM bank. With this layout, each bank can locally generate a complete portion of the output.

Most prior studies on accelerating LLM inference with PIM do not explicitly consider the memory attribute and weight layout inconsistencies between the prefill and decode phases (see Table~\ref{pim-llm-work}). IANUS~\cite{seo2024ianus}, AttAcc~\cite{park2024attacc}, NeuPIMs~\cite{heo2024neupims}, and AiM~\cite{kim2024skhynix_aimx} primarily focus on architectural optimizations, KV-cache acceleration, and scheduling techniques. An exception is that HBM-PIM~\cite{kim2024breakthrough} and PAISE~\cite{lee2025paise} circumvent the memory attribute inconsistency by placing host-friendly weights for prefill in a cacheable region and PIM-aware weights for decode in a non-cacheable region~\cite{pimlibrary, pimaicompiler}. However, such weight duplication introduces substantial DRAM capacity overhead.

\begin{table}[t]
\caption{PIM related works to accelerate LLM}
\label{pim-llm-work}
\vskip 0.1in
\begin{center}
\begin{small}
\begin{sc}
\begin{tabular}{l >{\centering\arraybackslash}p{1.7cm} >{\centering\arraybackslash}p{1.4cm} >{\centering\arraybackslash}p{1.4cm}}
\toprule
Works & Memory attribute & Weight layout & SW-only solution \\
\midrule
IANUS & $\times$ & $\times$ & $\times$ \\
AttAcc & $\times$ & $\times$ & $\times$ \\
NeuPIMs & $\times$ & $\times$ & $\times$ \\
AiM & $\times$ & $\times$ & $\times$ \\
HBM-PIM & $\triangle$ & $\triangle$ & $\times$ \\
PAISE & $\triangle$ & $\triangle$ & $\bigcirc$ \\
FACIL & $\times$ & $\bigcirc$ & $\times$ \\
PIM-SHERPA & $\bigcirc$ & $\bigcirc$ & $\bigcirc$ \\
\bottomrule
\end{tabular}
\end{sc}
\end{small}
\end{center}
\vskip -0.2in
\end{table}

\subsection{Current solutions and limitations}
\label{ssec:dup-re}

\label{ssec:weight-dup}
\begin{figure}[!tb]
  \center
  \includegraphics[width=0.95\columnwidth]{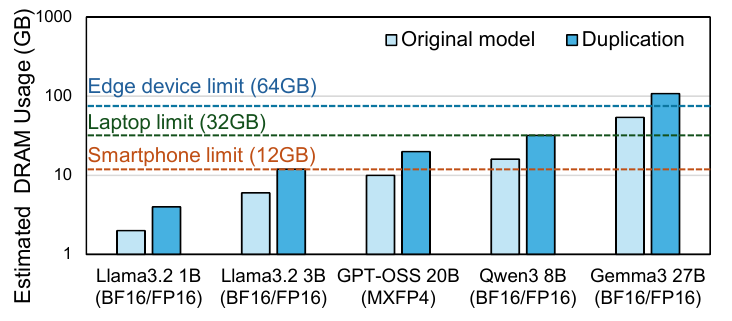}\caption{
    Estimated DRAM usage before and after weight duplication across representative LLMs.
    Y-axes use a logarithmic scale.
  }
  \label{fig:weight-dup}
  \vspace{-0.15in}
\end{figure}

\begin{table*}[t]
\caption{Latency modeling across input sequence lengths (Samsung Galaxy S24+ CPU, FLOP/B $\approx$ 4)
}
\label{tab:owr-overhead}
\vskip 0.1in
\centering
\begin{small}
\begin{sc}
\begin{tabular}{
  L{0.9cm}   
  C{0.8cm}   
  C{0.8cm}   
  C{1.0cm}   
  C{1.0cm}   
  C{1.6cm}   
  C{1.0cm}   
  C{1.6cm}   
}
\toprule
Input& GEMM& DRAM& Online&
\multicolumn{2}{c}{SUM(GEMM, Online)} &
\multicolumn{2}{c}{MAX(GEMM, Online)} \\
\cmidrule(lr){5-6} \cmidrule(lr){7-8}
& & & & time & vs GEMM & time & vs GEMM \\
\midrule
$1$ - $4$ & $t$  & $t$ & $3t$ & $4t$  & $400\%$ & $3t$  & $300\%$ \\
$8$      & $2t$  & $t$ & $3t$ & $5t$  & $250\%$ & $3t$  & $150\%$ \\
$16$     & $4t$  & $t$ & $3t$ & $7t$  & $175\%$ & $4t$  & $100\%$ \\
$32$     & $8t$  & $t$ & $3t$ & $11t$ & $138\%$ & $8t$  & $100\%$ \\
$64$     & $16t$ & $t$ & $3t$ & $19t$ & $119\%$ & $16t$ & $100\%$ \\
$128$    & $32t$ & $t$ & $3t$ & $35t$ & $109\%$ & $32t$ & $100\%$ \\
$192$    & $48t$ & $t$ & $3t$ & $51t$ & $106\%$ & $48t$ & $100\%$ \\
\bottomrule
\end{tabular}
\end{sc}
\end{small}
\vskip -0.15in
\end{table*}

\textbf{Weight duplication:} A straightforward approach to maintain performance across both phases is to keep two separate copies of the model weights, each tailored to the specific memory attributes and access patterns of the corresponding phase. However, the weight duplication worsens DRAM capacity constraints, particularly in LPDDR-based mobile and edge environments. Figure~\ref{fig:weight-dup} shows the estimated DRAM usage under duplication. Llama 3.2 3B requires at least 6 GB of DRAM capacity. Because modern smartphones typically provide 8 - 12 GB DRAM capacity, smartphones can execute the Llama 3.2 3B model without duplication. However, enabling PIM requires an additional 6 GB for a second copy, which makes deployment infeasible on smartphones. Similarly, Qwen3 8B and Gemma3 27B can run on laptops or edge devices without duplication, but they become infeasible once PIM is enabled. As a result, adopting PIM may force the use of smaller, lower accuracy models.

\textbf{Weight rearrangement:} Without duplication, both host and PIM can be supported by converting the PIM-aware weight layout to a host-friendly layout at runtime, immediately before GEMM. However, online weight rearrangement can incur significant latency overhead due to additional resource usage. FACIL~\cite{seo2025facil} reports that, for relatively short input SLs, online rearrangement increases the time-to-first-token (TTFT) by a factor of three.

To mitigate this overhead, FACIL augments the memory controller with lightweight address translation logic, which flexibly maps physical to DRAM addresses. This enables the elimination of the need for weight duplication and online rearrangement. FACIL maintains a single PIM-aware copy of the weights. During the prefill phase, the added logic presents the PIM-aware layout to the host in a host-friendly form, thereby preserving the GEMM performance on the host. However, the approach assumes a modified memory controller, which limits applicability in systems where the controller is fixed or cannot be changed. In addition, FACIL does not consider the inconsistency in memory attribute between the phases, which remains a fundamental challenge for PIM-enabled LLM deployment.

\subsection{Revisiting online weight rearrangement overheads}
\label{ssec:revisit}

FACIL deems online weight rearrangement impractical due to its substantial latency overhead. However, as the input SL continues to increase, the static cost of online rearrangement can be expected to occupy a smaller fraction of end-to-end latency. Especially, the spread of interactive LLMs~\cite{jeong2025flashgen} and on-device RAG~\cite{seemakhupt2024edgerag, park2025mobilerag} continues to enlarge input SL. In interactive LLMs, the response from the previous decode phase is appended to the next user query. Similarly, on-device RAG constructs inputs by concatenating user queries with context retrieved from a database, improving accuracy under constrained model sizes. Motivated by these trends, this work revisits the conditions under which online weight rearrangement becomes practical and quantifies its overhead as a function of input SL.

The overhead of the online weight rearrangement depends on the arithmetic intensity ($FLOP/B$) and the input SL. Table~\ref{tab:owr-overhead} models the overhead on a Samsung Galaxy S24+ device and shows how the overhead scales with varying input SL. For small inputs ($SL$ $<$ $4$), GEMM latency is normalized to a base time ($t$). As input SL increases beyond four, latency grows roughly proportionally ($(input$ $SL / (FLOPS/B))$ $\times$ $t$). The cost of online weight rearrangement is modeled as three DRAM transactions ($3t$): a read from the non-cacheable weight, a read of the destination from DRAM to cache, and a write to DRAM. $SUM(GEMM, Online)$ denotes serial execution, whereas $MAX(GEMM, Online)$ denotes overlapped execution. For input SL 1 - 8, fixed rearrangement cost dominates, resulting in 1.5 - 4$\times$ higher latency than GEMM alone. For input SL 16 - 192, growing GEMM time allows hiding rearrangement under computation, as $MAX(GEMM, Online)$ indicates. Above input SL 192, GEMM dominates, so the rearrangement overhead shrinks.

These observations indicate that for real workloads with sufficiently large input SL, the relative cost of online weight rearrangement can be amortized through the overlap with GEMM. Motivated by this insight, we propose PIM-SHERPA, a software-only method optimized for large input SL.
\section{Contribution}
\label{sec:contribution}

\begin{figure*}[!tb]
  \center
  \includegraphics[width=\textwidth]{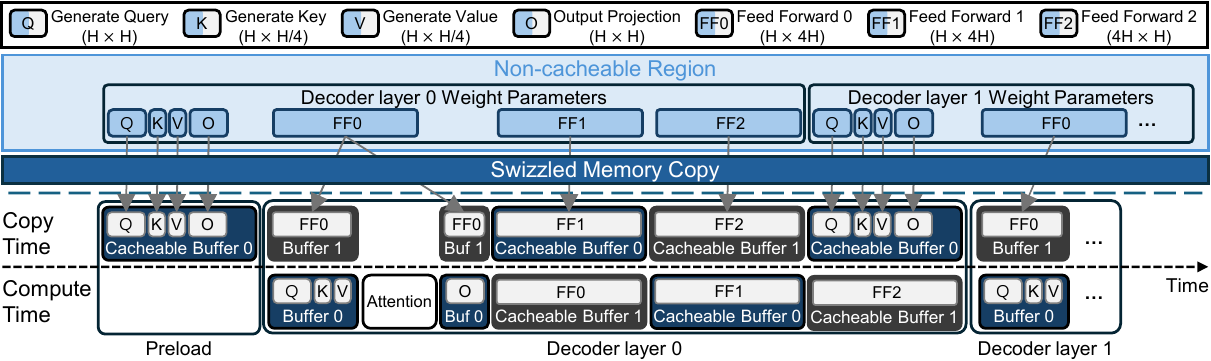}\caption{
    Illustration of DRAM double buffering. Two small DRAM buffers are allocated in the cacheable region to prefetch weights of the next layer from the non-cacheable region while GEMM for the current layer executes. This double-buffering mechanism effectively hides the latency of online rearrangement behind compute time. Each decoder layer (Q, K, V, O, FF0 (gate), FF1 (up), and FF2 (down)) corresponds to the Llama 3.2 structure, where H denotes the hidden size.
  }
  \label{fig:ddb}
  \vspace{-0.07in}
\end{figure*}

\begin{figure}[!tb]
  \center
  \includegraphics[width=0.9\columnwidth]{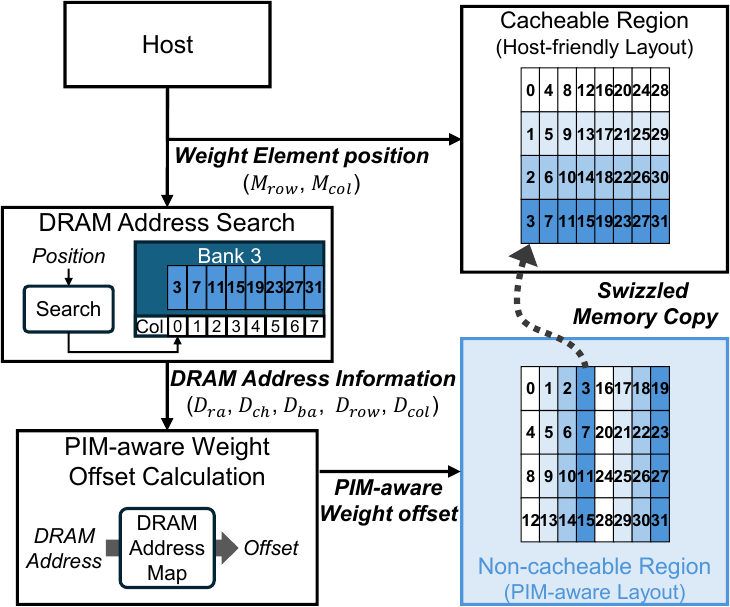}
  \caption{
    Swizzled memory copy process
  }
  \label{fig:smc}
  \vspace{-0.07in}
\end{figure}



To address the inconsistency in memory attribute and weight layout between prefill and decode phases, we introduce PIM-SHERPA, a software-only method for on-device LLM inference. The Key idea is to allocate small DRAM buffers in the cacheable region and dynamically relocate weights from non-cacheable region into these buffer at runtime.

\subsection{DRAM double buffering}
\label{ssec:ddb}

To hide the latency of the online weight rearrangement, we first employ DRAM Double Buffering (DDB), which overlaps data movement with computation through a double buffering scheme.
The cacheable DRAM buffer size is set to twice the weight size of the feed forward layer, which is typically the largest weight in a decoder with a uniform, repeated structure. During the prefill phase, while one cacheable buffer (buffer 0) is used to execute GEMM for the current layer, the weights for the next layer are prefetched from the non-cacheable region into the other buffer (buffer 1). The buffers alternate across successive decoder layers, allowing computation and data transfer to proceed in parallel. The proposed DDB eliminates the memory attribute inconsistency and performs weight rearrangement at runtime without requiring offline model duplication.

Figure~\ref{fig:ddb} illustrates how DDB is applied to the Llama 3.2 structure. The query (Q) and output projection (O) layers have size \textit{H$\times$H}. Because Llama 3.2 employs grouped-query attention~\cite{ainslie2023gqa}, the generate key (K) and generate value (V) layers are one quarter the size of the Q layer. The feed forward (FF) layer is commonly four times larger than Q layer, although recent models may deviate from this ratio. The cacheable DRAM buffer size is set to \textit{2$\times$H$\times$4H}, which equals twice the FF layer. All weight parameters reside in the non-cacheable region.

Before prefill begins, a preload step copies the GEMM weights into cacheable buffer 0. Because buffer 0 has size \textit{H$\times$4H}, it can hold the weights of the Q, K, V, and O layers. To balance computation and transfer overhead, weights of these four layers are grouped and stored in buffer 0. When GEMM operations for Q, K, and V layers are executed, the weight of FF0 layer is copied into buffer 1 at the same time. This divide the weight of FF0 into four equal parts for load balancing so that one quarter is transferred during each of the three layers. Because DDB hides memory copies under compute, swizzled memory copies are not issued during relatively memory bound layers such as non-linear functions, normalization, and attention. As shown in the figure, the remaining quarter of the FF0 copy is completed during the O layer. FF0, FF1, and FF2 layers have a weight size equal to the DRAM buffer capacity and therefore fully occupy the buffer. During FF0 and FF1 layers, the amount of weight copied matches the amount consumed by the ongoing computation. FF2 layer performs a preload for the next decoder layer by preparing the weights of the Q, K, V, and O layers, which minimizes pipeline bubbles. The online weight rearrangement time of FF2 layer is the smallest because K and V layers have one quarter size of weight and the total weight size of Q, K, V, and O is \textit{H$\times$2.5H}.

\subsection{Swizzled memory copy}
\label{ssec:smc}

Swizzled memory copy (SMC) is a key technique for resolving both memory-attribute and weight-layout inconsistencies. There are two approaches: (1) maintain the PIM-aware layout while copying from the non-cacheable to the cacheable region and incorporate swizzling logic into the GEMM kernel, or (2) convert the weights into a host-friendly layout during the copy so that the standard GEMM kernel can execute without modification. The first approach requires backend modifications and is hard to deploy with closed-source libraries, whereas the second confines the conversion to the copy path and leaves the backend untouched. Prioritizing portability and ease of integration, we adopt the second option and perform GEMM immediately after SMC.

\begin{figure}[!tb]
  \center
  \includegraphics[width=\columnwidth]{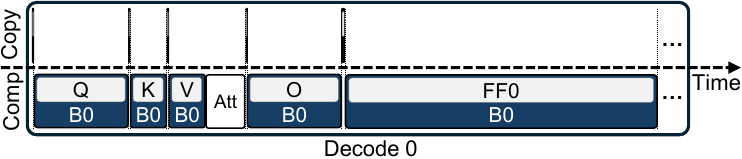}\caption{
    Illustration of online weight rearrangement with swizzled memory copy
  }
  \label{fig:owr}
  \vspace{-0.07in}
\end{figure}

Figure~\ref{fig:smc} illustrates the SMC process. To perform SMC, the host selects the element coordinates (\(M_{row}\), \(M_{column}\)) of the weight matrix to copy. These coordinates are used to locate the weight vector in DRAM that contains the (\(M_{row}\), \(M_{column}\)) element. For example, if the host specifies the position of the third vector, the mapping indicates DRAM bank 3 and DRAM row 0. The DRAM address information is applied to the address mapping to compute the PIM-aware weight offset. During the copy, the element coordinates determine the destination and the PIM-aware offset serves as the source, and the memory copy moves data from the non-cacheable region to the cacheable region.

\subsection{Online rearrangement with swizzled memory copy}
\label{ssec:owr}

PIM-SHERPA also offers online weight rearrangement with SMC (OWR), which performs the swizzled memory copy serially just before GEMM when the input SL is sufficiently long and GEMM time dominates. OWR uses a single cacheable buffer that holds only one layer of weights required by the current computation step. The buffer size is configured as \textit{H$\times$4H}, corresponding to the FF layer weight size, which is typically the largest layer in the model. 

As illustrated in Figure~\ref{fig:owr}, once required weights are copied into the buffer, OWR triggers GEMM using the rearranged data. This serialized procedure ensures a deterministic execution order and uniform memory access behavior across layers. This design can improve software simplicity by eliminating the need for complex interleaving or sophisticated synchronization between SMC and compute phases.
OWR is particularly effective in long-sequence scenarios where computation dominates overall performance. Because the SMC latency remains static regardless of the input SL, its relative impact on total runtime diminishes as the GEMM time increases. 

\subsection{Implementation details}
\label{ssec:implement}

We use ExecuTorch~\cite{executorch} as the base runtime for on-device LLM. When it comes to DDB, we deploy compute and copy threads; copy threads run SMC, loading weights from the non-cacheable region, swizzling the PIM-friendly layout to a host-friendly layout, and storing the result in the cacheable region. In parallel, compute threads execute GEMM with the optimized library supplied by the framework, without modification. Compute threads read tiled weights from the cacheable buffer not used by the copy thread, and two buffers in parallel overlap data movement with computation to hide rearrangement latency. DDB synchronizes the two thread groups once per layer. OWR creates and joins both copy and compute threads at each step. Multiple cores first perform SMC into a single buffer, and the host then runs GEMM on the prepared weights. Different from DDB, OWR requires no explicit inter-thread synchronization, which lowers implementation complexity.
\section{Experimental Setup}
\label{sec:experimetnal_setup}


We evaluated PIM-SHERPA using ExecuTorch~\cite{executorch}. We used a Samsung Galaxy S24+ and BF16 data format, as this configuration is officially supported to run Llama models on the device. The Samsung Galaxy S24+ is equipped with Exynos 2400 (see Table~\ref{tab:hw_pim_spec}) and its CPU system consists of a single big core (Cortex-X4), two middle-high and three middle-low cores (A720), and four little cores (A520). In order to utilize Neon SIMD architecture, we only used big and middle cores.

The six CPU cores together achieve a peak performance of 321 GFLOPS. The system has LPDDR5X-8533 with four channels, providing up to 68.264 GB/s, and FLOP/B is 4.7. We used the Llama 3.2 1B and 3B models released as an on-device model by META~\cite{meta2024llama32_blog}. The hidden sizes for Llama 3.2 1B and 3B models are $2048$ and $3072$, respectively, and the batch size was set to $1$. We assigned four CPU cores to GEMM because using six cores introduces tail effects that prevented full utilization of Neon SIMD, yielding performance similar to achieved with four cores. Two cores were allocated for copy threads.

\begin{table}[t]
\caption{Host and PIM emulation system specification}
\label{tab:hw_pim_spec}
\vskip 0.15in
\centering
\begin{small}
\begin{sc}
\begin{tabularx}{\columnwidth}{L{1.1cm} L{2.2cm} Y}
\toprule
Section & Field & Value \\
\midrule
\multirow{5}{1.1cm}{Host System}
  & Device & Samsung Galaxy S24+\\
  & Host CPU & 1 $\times$ Cortex-X4 ~~~~~~~~~~~~~~~ 2 $\times$ A720 @ 2.9GHz ~~~~~~~~~~~ 3 $\times$ A720 @ 2.7GHz \\
  & BF16 GFLOPS & 321  \\
  & DRAM BW & 68.264 GB/s \\
\midrule
\multirow{4}{1.1cm}{PIM Emulation System}
  & Device & LPDDR5X\mbox{-}PIM \\ 
  & Input RF  & 8 \\
  & Output RF  & 8 \\
  & ALU precision & BF16 \\
\bottomrule
\end{tabularx}
\end{sc}
\end{small}
\vskip -0.1in
\end{table}

\begin{figure}[!tb]
  \center
  \includegraphics[width=\columnwidth]{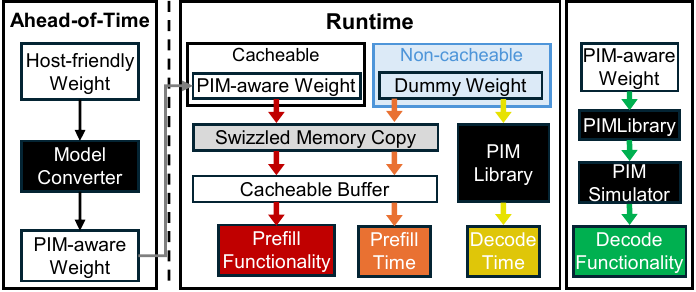}\caption{
    Prefill-decode execution time and functionality evaluation pipeline. Red and green indicate functionality paths for prefill and decode, respectively. Orange and yellow indicate execution time paths. Black boxes denote supported softwares. 
  }
  \label{fig:experiment}
  \vspace{-0.15in}
\end{figure}


We describe the execution time and functionality paths shown in Figure~\ref{fig:experiment} in detail. We used a model converter to transform host-friendly weights into PIM-aware weights based on the address mapping of the Galaxy S24+. The PIM execution model requires placing the weights in a single physically contiguous memory region and mapping it as non-cacheable to support direct device access. In Linux, such large physically contiguous allocations are typically provided by the contiguous memory allocator (CMA). 

\begin{table}[t]
\caption{GEMV execution time comparison between HBM-PIM and HBM-PIM emulation}
\label{tab:hbm_pim_emu_error}
\vskip 0.15in
\centering
\begin{small}
\begin{sc}
\begin{tabularx}{\columnwidth}{L{0.9cm} L{0.9cm} L{1.5cm} L{1.5cm} L{1.5cm}}
\toprule
Output Dim & Input Dim & HBM-PIM Relative Time & Emulation Relative Time & Emulation Error \\
\midrule
\multirow{4}{1cm}{4096}
  & 2048  & 1 & 0.965 & 3.54\% \\
  & 4096  & 1 & 0.980 & 2.02\% \\
  & 8192  & 1 & 0.969& 3.05\% \\
  & 16384 & 1 & 0.993 & 0.66\% \\
\midrule
\multirow{4}{1cm}{8192}
  & 2048  & 1 & 0.990 & 0.97\% \\
  & 4096  & 1 & 0.993 & 0.69\% \\
  & 8192  & 1 & 0.999 & 0.14\% \\
  & 16384 & 1 & 0.996 & 0.35\% \\
\bottomrule
\end{tabularx}
\end{sc}
\end{small}
\vskip -0.1in
\end{table}

\textbf{Prefill time emulation} (orange): SMC converts PIM-aware weights from a non-cacheable region and transfers resulting host-friendly weights to a cacheable buffer. In parallel, the host performs GEMM using the host-friendly weights stored in the cacheable buffer. We measured the prefill time while SMC and GEMM run concurrently. On the Galaxy S24+, the CMA-backed contiguous region that can be configured as non-cacheable is approximately 1~GB, while the Llama~3.2~1B and 3B models require at least 2.4~GB and 6~GB of weights, respectively. The device cannot accommodate all weights within the available non-cacheable contiguous region. Therefore, we used dummy weights for prefill time emulation. If the full LLM weights could fit in the CMA-backed contiguous region, the same PIM-aware weights could be used for both execution time and functionality paths, enabling measurement of actual execution time on the device rather than emulation. 

\textbf{Prefill functionality validation} (red): To verify functionality, we needed to use real PIM-aware weight. However, as previously mentioned, entire weights cannot be stored in the non-cacheable region. Thus, we placed the weights into the cacheable region and performed SMC, where PIM-aware weights were converted into a host-friendly layout within the cacheable buffer. We then performed GEMM using this layout and validated the functionality by comparing the prefill outputs of PIM-SHERPA against those of the original ExecuTorch.

\textbf{Decode time emulation} (yellow): We emulated an LPDDR5X-PIM system using LPDDR5X and PIMLibrary. We modified PIMLibrary to support ARM CPUs and emulated LPDDR5X-PIM performance. To ensure that every PIM request reaches the memory controller, we use the dummy weight in the non-cacheable region. We measured the execution time of LPDDR5-PIM using PIMLibrary with the dummy weight.

To substantiate the accuracy of our PIM emulation, we compare real HBM-PIM hardware against HBM-PIM emulation and observe a small performance gap. Table~\ref{tab:hbm_pim_emu_error} reports GEMV execution times measured on real HBM-PIM hardware and on HBM-PIM emulation. Across multiple input and output dimensions, the emulation error remains low (0.1\% - 3.6\%) relative to real HBM-PIM. The small gap can be explained as follows. HBM-PIM does not introduce additional PIM-specific requests; instead, it operates using conventional DRAM read and write requests. Moreover, HBM-PIM uses the same memory controller and timing parameters as conventional HBM. Therefore, when we issue the same DRAM request pattern that triggers HBM-PIM GEMV execution on an AMD MI100 GPU system equipped with conventional HBM, the externally observed memory access and service time closely matches that of real HBM-PIM. We incorporated LPDDR5X-PIM parameters into the HBM-PIM emulation system to develop a high-fidelity LPDDR5X-PIM emulation system, which has undergone industry-level validation.

\textbf{Decode functionality validation} (green): Without an actual PIM device, we cannot directly verify whether the emulation produces correct GEMV results. To address this, we validate functionality by coupling PIMLibrary with a PIM Simulator. The simulator performs GEMV using real input and weight values. For validation, we store LLM weights in the simulator DRAM at the same DRAM addresses used on the mobile device. We then feed the DRAM requests generated by PIMLibrary into the simulator and compare the resulting GEMV outputs against those computed by the host CPU.
\section{Evaluation}
\label{sec:evaluation}

\begin{figure}[!tb]
  \center
  \includegraphics[width=0.9\columnwidth]{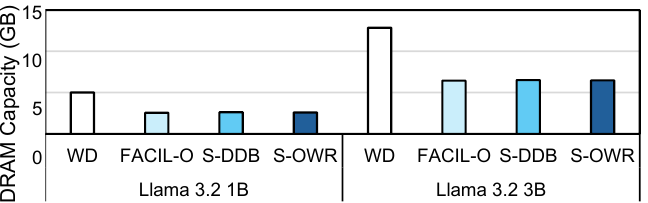}\caption{
    Required DRAM capacity across WD, FACIL-O, S-DDB and S-OWR
  }
  \label{fig:eval1}
  \vspace{-0.15in}
\end{figure}

\begin{figure}[!tb]
  \center
  \includegraphics[width=\columnwidth]{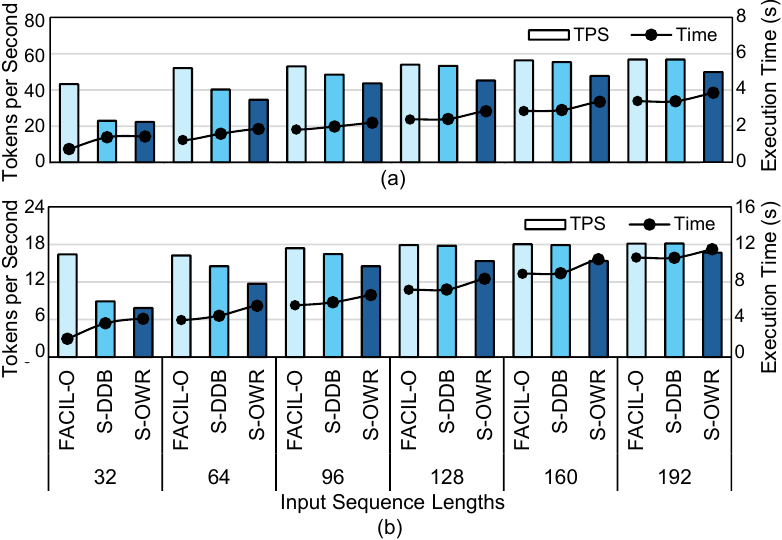}\caption{
    TTFT across input sequence lengths using Llama 3.2 (a) 1B and (b) 3B models. Results are presented by tokens per second (TPS) and execution time. When it comes to TTFT, the result of WD is the same as the FACIL-O.
  }
  \label{fig:eval2}
  \vspace{-0.15in}
\end{figure}

\begin{figure*}[!tb]
  \center
  \includegraphics[width=\textwidth]{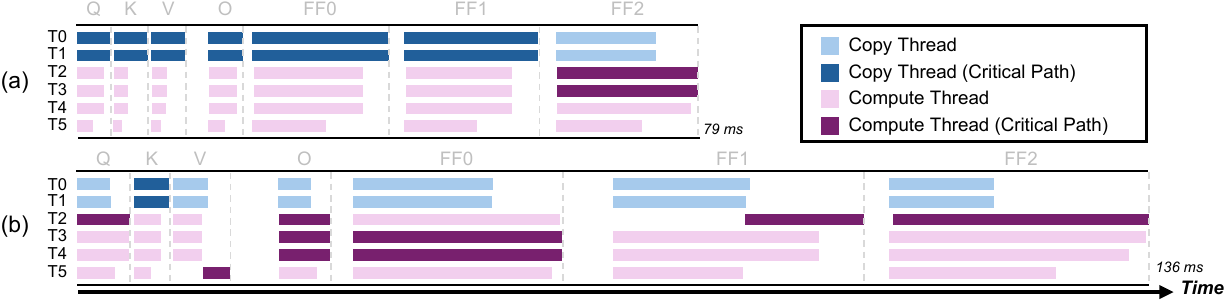}\caption{
    S-DDB execution timeline of a single decoder layer at input sequence lengths of (a) 64 and (b) 128 tokens on the Galaxy S24+ using Llama 3.2 1B model. Blue bars denote the copy thread and purple bars indicate the compute thread. Each bold color bar highlights the critical path within each layer. Thread 5 is the main thread and is bound to the Cortex-X4. Threads 0 and 1 are pinned to the A720 cores with 2.9 GHz and threads 2 - 4 are pinned to the A720 cores with 2.7 GHz.
  }
  \label{fig:eval3}
  \vspace{-0.15in}
\end{figure*}

We present evaluation results by applying PIM-SHERPA to Llama 3.2 inference. To understand the results in detail, we consider four scenarios: weight duplication (WD), FACIL under an oracle assumption (FACIL-O), SHERPA-DDB (S-DDB), and SHERPA-OWR (S-OWR). In WD, host-friendly weights are stored in a cacheable region, whereas PIM-aware weights are placed in a non-cacheable region. To compare FACIL with PIM-SHERPA, we assume that FACIL resolves a memory attribute inconsistency; we denote this as FACIL-O.

\textbf{DRAM capacity saving:}
S-DDB and S-OWR markedly reduce the required DRAM capacity, as shown in Figure~\ref{fig:eval1}. For the Llama 3.2 1B BF16 model, host-friendly weights require 2.47~GB.
To avoid misalignment, PIM-aware weights include additional padding in the K and V layers and in the LM head. This padding increases size by 80~MB relative to the host-friendly baseline. In WD, both host-friendly and PIM-aware weights are stored, whereas S-DDB and S-OWR only require PIM-aware weights and small additional cacheable buffers. This results in memory savings of 47.8\% and 48.5\% relative to WD. The cacheable buffer size is 32~MB (2048$\times$8192$\times$2~B). Because S-OWR uses a single cacheable buffer, its required capacity is about 0.7\% smaller than S-DDB. FACIL-O augments the memory controller and keeps only PIM-aware weights, reducing the required capacity by 49.2\%. The Llama 3.2 3B BF16 model requires 6.4~GB for host-friendly weights, and S-DDB and S-OWR save 49.4\% and 49.7\% of DRAM capacity, respectively. With software-only supports, S-OWR and S-DDB achieve significant capacity savings comparable to FACIL-O while resolving the memory attribute and weight layout inconsistencies.

\textbf{TTFT performance: }
The TTFT of PIM-SHERPA is nearly identical to, or slightly lower than, that of FACIL-O as the input SL increases. Figure~\ref{fig:eval2} shows that S-DDB achieves TTFT comparable to FACIL-O for the input SL between 128 and 192. S-DDB uses two copy threads, and the static SMC latency from the non-cacheable region to the cacheable region is about 0.89 and 2.54 s on Llama 3.2 1B and 3B models, respectively. When the GEMM compute time exceeds this latency (SL $\ge$ 128), S-DDB effectively hides the SMC latency. In addition, as the input SL increases, the performance of S-OWR gradually approaches that of the others. At 192-input SL on the 3B model, S-OWR achieves 16.7 TPS, nearly matching FACIL-O and S-DDB. S-OWR performs the SMC and GEMM sequentially using four threads. The measured SMC latencies are approximately 0.6 s (1B) and 1.4 s (3B), respectively. As a result, S-OWR always incurs SMC latency and lags behind the others.

The overlap achieved by S-DDB differs from the theoretical estimate in Section~\ref{ssec:revisit}. Table~\ref{tab:owr-overhead} indicates that the SMC latency of S-DDB should overlap with GEMM time at the 16-input SL. However, we found that data movement from the non-cacheable to the cacheable regions is about twice as slow as movement between cacheable regions through experiments. In addition, using only two copy threads does not saturate available memory bandwidth. As a result, SMC achieves only about one quarter of the peak DRAM bandwidth. At 32-input SL on the 1B model, S-DDB and S-OWR exhibit similar latencies: S-OWR combines sums 0.74 s of GEMM time with 0.6 s of SMC latency on four threads, whereas S-DDB incurs about 0.89 s with two copy threads plus additional contention overhead.

\textbf{S-DDB analysis:}
Figure~\ref{fig:eval3} shows the per-thread timeline within a decoder layer. The copy threads perform SMC once for PIM-aware weights regardless of input SL, resulting in a constant SMC latency. In contrast, GEMM time scales linearly with input SL. At 64-input SL, the copy thread becomes the critical path for most layers (Figure~\ref{fig:eval3}(a)). Because K and V layers use grouped-query attention, their matrices are smaller, which makes their compute time much shorter than that of other layers. Because the K and V layers copy the same amount of FF0 weight as the Q and O layers, the gap between compute time and copy latency becomes pronounced. The FF2 layer prefetches the weights of Q, K, V, and O layer, partially hiding the SMC latency because the weights of K and V layer are relatively small.

At 128-input SL, the SMC latency of all layers except K can be hidden to the compute latency (see Figure~\ref{fig:eval3}(b)). In the V layer, even though the compute time is shorter, the layer may complete later due to scheduling issues when hardware threads are unavailable. Because the number of SIMD units on a Coretex-X4 is twice that of an A720, T5 finishes GEMM earlier across all layers. In the FF1 layer, T2 fails to enqueue into the run queue due to limited scheduling slots and remains pending until T5 completes on the Coretex-X4. Once T5 completes, the scheduler dispatches T2 to the available core, resulting in delayed completion.

\begin{figure}[!tb]
  \center
  \includegraphics[width=\columnwidth]{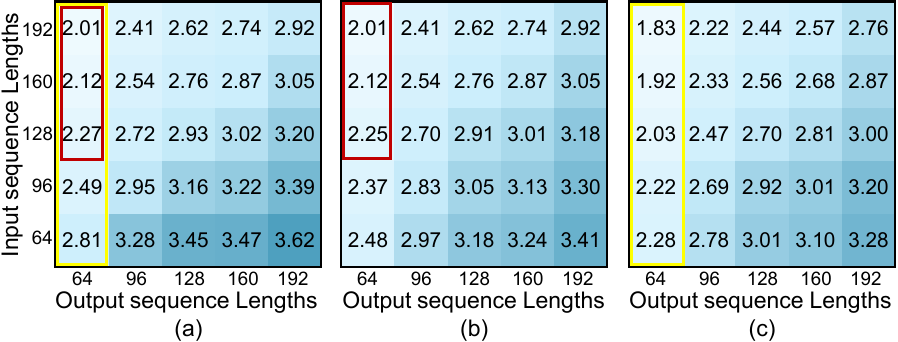}\caption{
    End-to-end inference speedup of (a) FACIL-O, (b) S-DDB, and (c) S-OWR over a host-only LPDDR5X system when using LPDDR5X-PIM. Input and output sequence lengths vary from 64 to 192 . We used Llama 3.2 1B model.
  }
  \label{fig:eval4}
  \vspace{-0.2in}
\end{figure}

\begin{figure}[!tb]
  \center
  \includegraphics[width=\columnwidth]{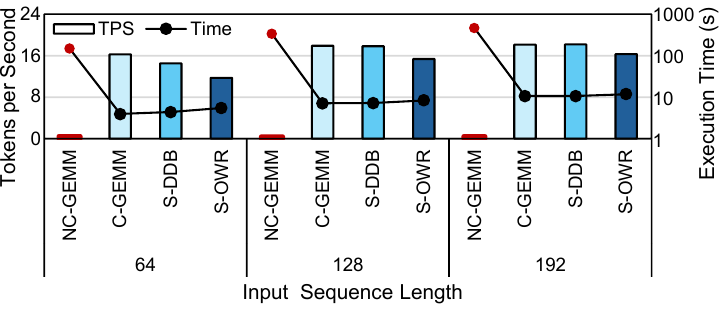}\caption{
    TTFT comparison when the device has only a single weight due to limited DRAM capacity. We used Llama 3.2 3B model. C-GEMM denotes GEMM with cacheable weight without PIM. NC-GEMM stands for GEMM with non-cacheable weights using PIM. Y-axes of execution time uses a logarithmic scale.
  }
  \label{fig:eval5}
  \vspace{-0.15in}
\end{figure}

\textbf{LLM inference speedup using PIM:}
S-DDB and FACIL-O exhibit similar end-to-end inference speedup with LPDDR5X-PIM across most combinations of input and output SLs (see Figure~\ref{fig:eval4}). For input SL with $\ge$ 128, the speedup of FACIL-O and S-DDB is nearly identical, even when output SL is relatively small (red box). As the input SL increases, the performance gap between S-OWR and the others gradually narrows (yellow box). For input SLs 64 - 96, S-DDB differs from FACIL-O by about 2.5\% - 19\%. Although S-OWR shows smaller gains, it still delivers up to 3.3$\times$ speedup over the non-PIM baseline.

\textbf{TTFT performance on limited DRAM capacity:}
PIM-SHERPA remains useful even when the host can keep only a single copy of the weights due to limited DRAM capacity (see Figure~\ref{fig:eval5}). To reflect realistic device constraints, we evaluated the Llama-3.2 3B model on the Galaxy S24+, which can hold only one copy of the weights. C-GEMM executes GEMM using weights in the cacheable region. During prefill, a C-GEMM can surpass S-DDB and S-OWR; however, once decode phase is included, it is not a viable substitute for PIM. Because NC-GEMM cannot reuse cached weights, it rereads weights from DRAM approximately input SL times. Across the evaluated range, the execution time of NC-GEMM grows roughly in proportion to input SL, reaching 148 s, 332 s, and 458 s at input SLs of 64, 128, and 192, respectively. Throughput remains about 0.4 TPS regardless of the input SL. These findings also confirm that the device correctly enforces the non-cacheable setting.
\section{Discussion}
\label{sec:disussion}
\textbf{Applicability to GPU and NPU systems:} PIM-SHERPA can be applied to GPU systems that include a last-level cache. The GPU tiles GEMM at thread-block granularity to distribute work across streaming multiprocessors (SMs)~\cite{li2019coordinated}. The GPU uses the L2 cache to avoid repeatedly reading the same data from DRAM across SMs. On a GPU with PIM, PIM-SHERPA can reduce data movement by approximately the input SL divided by the thread-block tile size during prefill. From an implementation perspective, it can leverage a GPU software pipeline~\cite{lym2019delta, lee2022interwarpmulticast}, which hides DRAM to shared memory transfers under compute. Integrating SMC into the pipeline preserves the same latency hiding benefit.

The application to NPUs depends on the memory hierarchy. If an NPU includes both on-chip SRAM and a cache~\cite{vasiljevic2024blackhole, coburn2025mtia2i}, the approach mirrors the CPU and GPU cases. If an NPU provides only SRAM~\cite{jouppi2023tpuv4, jia2019ipu}, we can allocate the double buffer directly in SRAM, and the system performs the swizzled memory copy from DRAM to SRAM, which avoids additional weight store traffic to DRAM. Thus, we emphasize the universality of our approach.

\textbf{Online rearrangement overhead over FLOP/B:} On systems with higher compute capability, the overhead of online rearrangement can appear larger. NVIDIA GPUs are representative of high FLOP/B platforms. NVIDIA Jetson AGX Orin (edge) and B100 (server) provide about 207 and 241 FLOP/B, respectively. With an input SL of 128, as used in our experiments, TTFT can drop to about one third of the theoretical maximum on such platforms. However, systems with high compute capability can also use the input SL more freely. Recent LLM work~\cite{hsieh2024ruler} reports effective input SLs up to 128K and, the growth in input SL can exceed the increase in FLOP/B by roughly an order of magnitude. Thus, the method remains applicable on high FLOP/B systems, although the preferred operating window in terms of FLOP/B and input SL may shift.

\section{Related Work}
\label{sec:related_work}
\textbf{PIM software works:} A body of research addresses PIM inefficiencies in software. SimplePIM~\cite{chen2023simplepim} simplifies programming with an iterator-based API and communication primitives. PIM-STM~\cite{lopes2024pimstm} shows that transactional memory applications can be effectively accelerated on PIM. PIM-CARE~\cite{hwang2025pimcare} provides a compiler for concurrent multi-application execution and significantly improves PIM resource utilization. These studies use the commercial UPMEM device~\cite{devaux2019truepim}, which positions DRAM as an accelerator and does not consider inconsistencies in memory attribute and weight layout. PAISE~\cite{lee2025paise} is a software study that uses Samsung HBM-PIM internally and improves PIM utilization with a data layout adjustment module. However, like HBM-PIM, it duplicates weights to avoid these inconsistencies.

\textbf{LLM inference acceleration using PIM:} Several studies accelerate LLM inference with PIM. IANUS~\cite{seo2024ianus} and NeuPIMs~\cite{heo2024neupims} orchestrate an NPU and PIM to execute layers with different characteristics in parallel. AttAcc~\cite{park2024attacc} partitions and balances requests with heterogeneous sequence lengths to accelerate the KV cache. These works do not consider memory attribute and weight layout inconsistencies. HBM-PIM~\cite{lee2021pim_isca} and PAISE~\cite{lee2025paise} circumvent these inconsistencies via weight duplication. FACIL~\cite{seo2025facil} resolves the weight layout inconsistency through flexible address mapping but does not consider the memory attribute inconsistency.

\section{Conclusion}
\label{sec:conclusion}

To the best of our knowledge, this is the first work to identify the memory attribute inconsistency and propose effective solutions on PIM enabled systems. PIM-SHERPA introduces two software only techniques that resolve both memory attribute and weight layout inconsistencies at runtime. On a commercial on-device platform, the approach reduces the DRAM capacity usage compared to weight duplication approximately 47.8 - 48.5\%. Our idea also delivers TTFT comparable to FACIL, while it still preserves PIM speedup. Our study suggests that addressing these challenges is crucial for the future commercialization of PIM.

\nocite{langley00}

\bibliography{ref}
\bibliographystyle{mlsys2025}



\end{document}